\documentclass[conference,10pt,a4paper]{IEEEtran}
\IEEEoverridecommandlockouts
\usepackage[T1]{fontenc}
\usepackage[utf8]{inputenc}
\usepackage[english]{babel}

\usepackage{amsmath,amssymb}
\usepackage{graphicx}
\usepackage{booktabs}
\usepackage{multirow}
\usepackage{array}
\usepackage{xcolor}
\usepackage{tikz}
\usetikzlibrary{positioning,arrows.meta,calc,fit,backgrounds,shapes.geometric,patterns}
\usepackage{pgfplots}
\pgfplotsset{compat=1.18}
\usepackage{url}
\usepackage{balance}

\usepackage[breaklinks=true]{hyperref}

\definecolor{qblue}{HTML}{2A78D6}
\definecolor{qgreen}{HTML}{008300}
\definecolor{qorange}{HTML}{EB6834}
\definecolor{qred}{HTML}{D03B3B}
\definecolor{qviolet}{HTML}{4A3AA7}

\newcommand{\gpk}{\ensuremath{gpk}}
\newcommand{\gsk}{\ensuremath{gsk}}
\newcommand{\usk}{\ensuremath{usk}}

\newcommand{\sigann}{\ensuremath{\sigma_{\mathrm{ann}}}}

\def\BibTeX{{\rm B\kern-.05em{\sc i\kern-.025em b}\kern-.08em T\kern-.1667em\lower.7ex\hbox{E}\kern-.125emX}}

\begin{document}

\title{Quantum-Seeded Distributed Post-Quantum Group\\ Signatures for 5G V2X Communications}
\author{
\IEEEauthorblockN{
P. Blanco-Casal,
P. Rejo-Iglesias,
D. Cajaraville-Aboy,
J. Vecino-Peñas,\\
R. P. Díaz-Redondo,
A. Fernández-Vilas,
and M. Fernández Veiga
}

\IEEEauthorblockA{
\textit{atlanTTic Research Center} 
\textit{Universidade de Vigo} 
Vigo, Spain
}
\thanks{
This work has been submitted to the IEEE for possible publication.
Copyright may be transferred without notice, after which this version may no longer be accessible.
}
}

\markboth{IEEE ICC 2026}%
{Blanco-Casal \MakeLowercase{\textit{et al.}}: Quantum-Seeded Distributed Post-Quantum Group Signatures for 5G V2X}

\maketitle


\begin{abstract}
  Vehicle-to-Everything (V2X) communications rely on elliptic-curve signatures that a quantum adversary breaks and on pseudonym certificates whose anonymity rests on rotation rate, not cryptography. This work presents qDGSP, a post-quantum V2X infrastructure: a hash-based dynamic group signature scheme (DGSP, over SPHINCS+ and WOTS+) distributed between a central controller and roadside verifiers. Group keys derive deterministically from quantum key distribution (QKD) seeds served over the ETSI GS QKD 014 interface, so no group-signature secret crosses the classical channel. qDGSP-EA (Ephemeral Association) binds an ephemeral ML-DSA-65 key to one group signature per 5G cell: messages remain unlinkable across cells, linkable within one. On a testbed with emulated 5G links, median end-to-end latency stayed within the 100\,ms ETSI budget for every variant. The qDGSP-EA intra-cell DENM matched the NIST ML-DSA-65 baseline on the wire, and under load EA mode raised delivery from roughly 78\% to 93\%.
\end{abstract}

\begin{IEEEkeywords}
Distributed verification, dynamic group signatures, ETSI ITS, hash-based signatures, PQC, QKD, V2X.
\end{IEEEkeywords}

\IEEEpeerreviewmaketitle

\section{Introduction}
\label{sec:intro}

\IEEEPARstart{V}{ehicles} exchange state and hazard information with each other (V2V), with roadside infrastructure (V2I) and with network services (V2N), collectively Vehicle-to-Everything (V2X) communication, over access networks converging on 5G and 6G~\cite{etsi_ts_122_186_v19,SANTA2022100537}. In Europe the ETSI Intelligent Transport Systems (ITS) family codifies the message formats, among them the Decentralized Environmental Notification Message (DENM)~\cite{etsi_en_302_637_3}, and the security profile protecting them~\cite{etsi_ts_103097_v211}. Cooperative ITS (C-ITS) thus inherits a standardised trust model: ECDSA over NIST P-256, with short-lived pseudonym certificates issued by a vehicular public-key infrastructure~\cite{etsi_ts_102_941}.

Shor's algorithm breaks the hardness assumption ECDSA rests on, and the exposure follows the \emph{harvest now, decrypt later} premise: an adversary who records V2X traffic today can later break the keys that protected it and retroactively strip the anonymity of vehicles. Recorded mobility does not expire, so migration to NIST-standardised post-quantum cryptography (PQC) is overdue for this traffic.

Beyond the quantum threat, the anonymity C-ITS offers today is heuristic. Nothing prevents two messages signed under the same pseudonym certificate from being linked; unlinkability rests on how often the certificate is rotated, so a passive observer can correlate messages in space and time and reconstruct a vehicle's trajectory~\cite{Sedar_V2X_security_survey}. Group signatures address this at the root: a verifier confirms that some group member signed without learning which one, while a designated authority can lift anonymity after an incident~\cite{perera_survey_groupring}. Because vehicles enter and leave continuously and an offender must stop being trusted immediately after detection, a fully dynamic construction is a requirement~\cite{dgsp2025}.

Deploying these primitives in a V2X network exposes three weaknesses that current 
systems do not address jointly. First, secret distribution: shipping group key material
over a classical channel compromises the trust root from the first 
exchange~\cite{dervisevic2024quantumkeydistributionnetworks}. Second, the integrity 
of anonymity: unlinkability is only as good as the entropy beneath it, and key material 
from low-entropy seeds or software pseudorandom generators leaves signatures
correlatable~\cite{perera_survey_groupring}. Third, quality of service: C-ITS hazard 
warnings run under a hard end-to-end budget of 100\,ms~\cite{etsi_tr_102_638,etsi_ts_122_186_v19} on a radio that fragments anything 
large, while a post-quantum group signature inflates the DENM by more than two orders
of magnitude over the ECDSA signature~\cite{techrxiv_pqc_v2x}. These weaknesses do not admit independent solutions: a PQC signature inside the existing vehicular PKI leaves unlinkability heuristic, a group signature distributed over a classical channel relocates the trust root onto the channel the adversary is already recording, and both inflate the wire format.

This paper presents qDGSP, which treats the three jointly. Its contributions are:
\begin{itemize}
  \item \textbf{A distributed post-quantum group signature seeded by a QKD network.} A controller and the verifiers derive the group key material locally and deterministically from high-entropy keys retrieved through the ETSI GS QKD 014 API~\cite{etsi_qkd_014}, so no secret of the scheme crosses the classical channel and groups are established on demand.
  \item \textbf{Integration into ETSI ITS critical messaging.} The scheme is bound to DENM traffic with a signed role container that drives role-based dissemination, and with an active semantic revocation cycle triggered by role-policy violations in the message content.
  \item \textbf{qDGSP-EA, an intra-cell amortisation mode.} Ephemeral Association binds a short-lived ML-DSA-65 key~\cite{nist_fips_204} to a single group signature for the duration of a vehicle's residence in one Roadside ITS-S cell, spreading the structural cost of the hash-based signature over every alert emitted there while keeping emissions unlinkable across cells.
  \item \textbf{A reproducible testbed and a quantitative evaluation.} A containerised Rust implementation,\footnote{\href{https://gitlab.com/Quantum-Safe-Architectures/qDGSP}{https://gitlab.com/Quantum-Safe-Architectures/qDGSP}} with emulated 5G delay and loss profiles and load levels dimensioned from a real ETSI ITS capture~\cite{v2aixmultimodalrealworlddataset}, measures signature size, CPU and provisioning cost, end-to-end latency, load behaviour and revocation propagation. 
\end{itemize}

\section{Preliminaries}
\label{sec:bg}

A group signature~\cite{GroupSignature} lets any member sign for the group: a 
verifier learns only that the signer is legitimate, while the manager can still 
open a signature to recover the identity. Three keys define the model~\cite{kaaniche2020privacy}: \gpk{} verifies, \usk{} signs, and the manager 
secret \gsk{} admits members and opens signatures. Fully dynamic schemes add 
\texttt{Join} and \texttt{Revoke} over a public list of members (PLM) and a revocation 
list (RL) without changing \gpk{}; vehicular use requires this, since a misbehaving 
vehicle must be excluded at run time.

DGSP~\cite{dgsp2025} is fully dynamic and hash-based. A single 
SPHINCS+~\cite{sphincs+} keypair belongs to the manager role and 
its hypertree root is \gpk{}, so verifying a group signature reduces to a SPHINCS+ 
verification. Users never sign with SPHINCS+: they generate WOTS+~\cite{wots} 
one-time keypairs in batches and have the manager certify the public keys, so 
signing consumes one unused key and attaches its certificate.
The certificate carries only an encrypted tag 
$\zeta_{\mathit{id},j} = E_k(\mathit{id} \,\|\, j)$, with $E$ a strong pseudorandom 
permutation under a manager-only key $k$ and $j$ the certificate index. Without $k$ 
the tag is indistinguishable from random, so a user's signatures are unlinkable. 
A companion commitment $\tau$, checked 
through a proof $\pi$, makes that opening publicly verifiable, so the manager cannot 
frame an innocent user nor the accused repudiate. Revocation acts on the public 
$\zeta$ tags already inside the signatures, a lookup that exposes no secret to verifiers.
Forward anonymity holds because tags are permutation outputs and WOTS+ signatures 
share no material.

QKD gives two endpoints identical symmetric keys with information-theoretic security, 
and~\cite{etsi_qkd_014} is the REST API through which authorised nodes retrieve the same 
key by its \texttt{key\_ID}; a request can name further authorised retrievers  (termed 
secure application entities (SAEs)) the multi-SAE authorisation qDGSP relies on. 
Two nodes holding the same QKD key derive identical hash-based material with nothing 
but the \texttt{key\_ID} crossing the classical channel.
qDGSP carries its signature material in ETSI DENM~\cite{etsi_en_302_637_3} rather 
than in periodic CAM traffic. For road-hazard warnings ETSI sets a 100\,ms end-to-end 
latency budget~\cite{etsi_tr_102_638,etsi_ts_122_186_v19}.

\begin{figure}
  \centering
  \includegraphics[width=0.99\columnwidth]{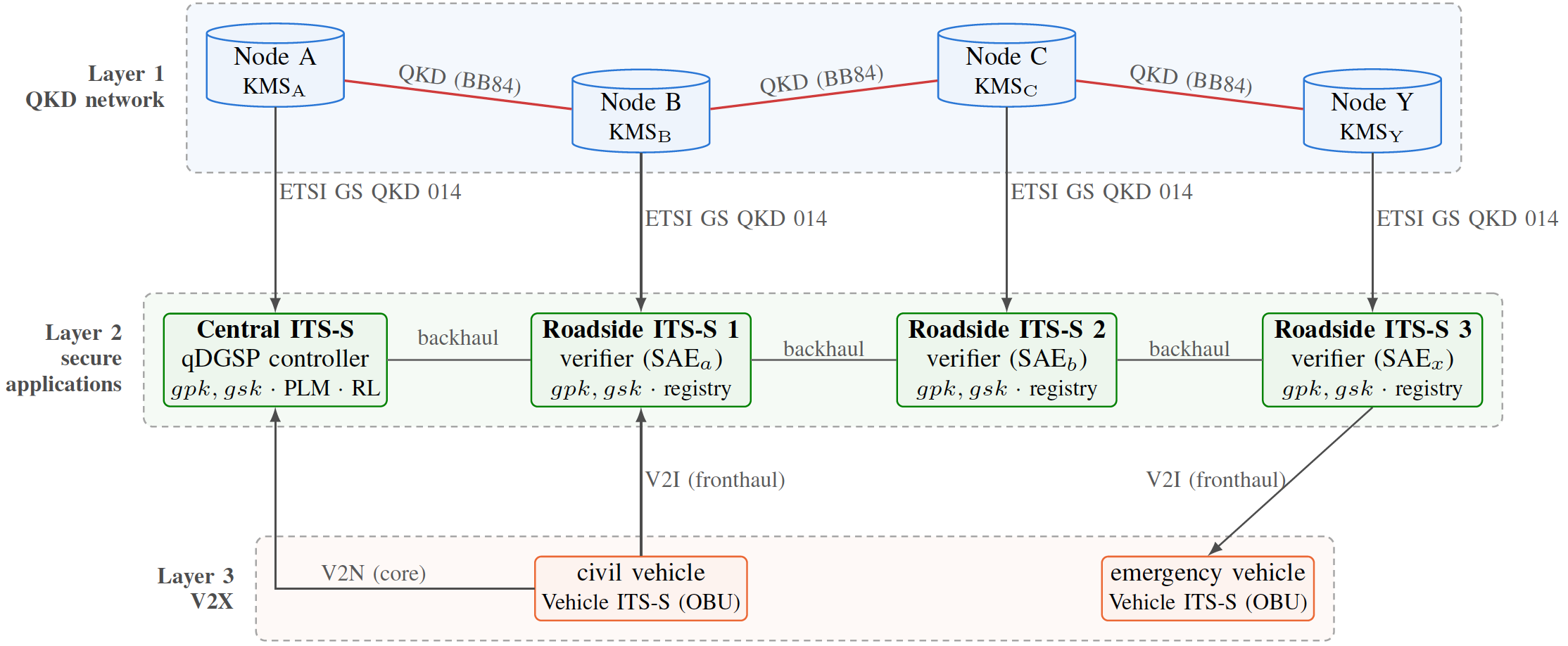}
  \caption{Three-layer qDGSP architecture. The QKD network delivers high-entropy 
  seed material through the ETSI GS QKD 014 interface; the controller and every 
  roadside verifier derive the same group key material locally from that seed.}
\label{fig:arch}
\end{figure}

\section{System Architecture}
\label{sec:arch}

qDGSP maps the scheme of Section~\ref{sec:bg} onto the ETSI ITS station 
model~\cite{ETSI_EN_302_665_V1.1.1}, targeting DENM rather than CAM traffic: 
DENMs carry the tightest end-to-end budget and the cause codes on which role-based 
routing and semantic revocation act. Fig.~\ref{fig:arch} shows three coupled layers 
of which DGSP spans  the upper two. Two principles organise the design. On-board units 
(OBUs) are grouped  by \emph{role} (e.g., civil traffic, emergency services) and the 
infrastructure routes each validated alert to the group concerned instead of flooding 
the area; and cryptographic lifetime is scoped to a \emph{cell}, the V2I coverage 
area of one Roadside ITS-S, to which qDGSP-EA binds its ephemeral signing material.

\subsection{Threat Model}
\label{subsec:threat}

A passive observer captures all V2X traffic in an area and correlates one vehicle's 
presence across zones, now or once a cryptographically relevant quantum computer exists.
Baseline mode renews the signing material for every message over quantum-resistant 
primitives; qDGSP-EA relaxes this to unlinkability between cells. Forward anonymity 
bounds the stored-traffic variant, but not the manager key, whose custody sustains 
anonymity against the infrastructure itself.
A compromised OBU emits a DENM with a \texttt{CauseCode} reserved for another 
functional group (role impersonation); the verifier detects this semantically and 
starts a revocation cycle that lifts the offender's anonymity at the controller.
A channel attacker induces loss or reordering, absorbed by the DENM repetition 
ETSI already foresees~\cite{etsi_en_302_637_3}.

Out of scope are physical attacks on OBU hardware, so the vehicle key store is 
assumed intact; active attacks on the QKD network, assumed trustworthy under the 
ETSI trusted-node model with an honest key management entity (KME) per node; 
and the wired backhaul, a trusted administrative domain.

\subsection{Entities}
\label{subsec:entities}

\subsubsection{QKD network and distributed key management system (DKMS)} A trusted 
KME at each infrastructure node, reachable through the ETSI GS QKD 014 
API~\cite{etsi_qkd_014}.

\subsubsection{Controller (Central ITS-S)} The logical authority: it performs 
the manager operations of Section~\ref{sec:bg} and maintains the authoritative 
PLM and RL, pushed to the verifiers.

\subsubsection{Verifiers (Roadside ITS-S)} The distributed validation plane: each 
verifier validates and routes messages as a delegated authority, without a real-time 
query to the controller, holding local copies of PLM and RL plus the volatile state 
introduced where it is used.

\subsubsection{Vehicle (Vehicle ITS-S)} Produces and consumes messages, registers 
with the controller and draws on a locally provisioned pool of WOTS+ certificates.

Four virtual subnetworks isolate traffic, one per transport segment. The 
fronthaul carries V2I traffic between OBUs and verifiers over UDP, matching the 
connection-less ETSI BTP-B~\cite{etsi_en_302636_5_1} and leaving per-frame loss 
and IP fragmentation exposed. The core models the 5G user plane and carries only 
registration and certificate requests, off the critical path. The backhaul 
interconnects verifiers and controller, carrying PLM and RL synchronisation, 
edge-raised revocation requests and inter-peer alert forwarding over HTTP/JSON. 
Finally, the QKD network reaches the DKMS nodes.

\subsection{Deterministic Seeding and On-Demand Group Setup}
\label{subsec:seeding}

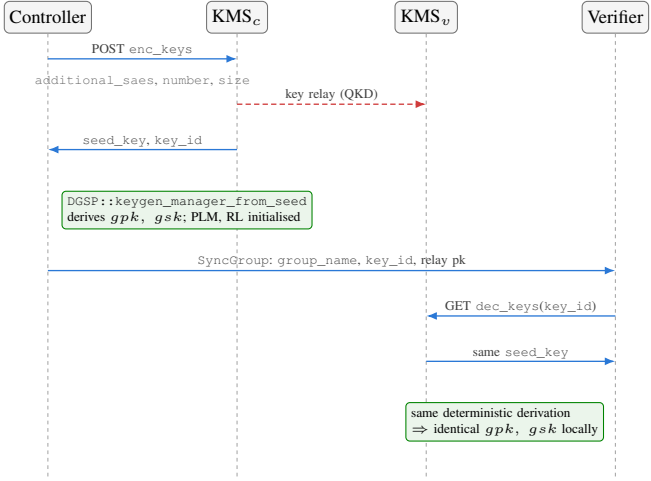
\begin{figure}[!t]
\centering
\begin{tikzpicture}[
  font=\scriptsize,
  actor/.style={rounded corners=2pt, draw=black!55, fill=black!4,
                inner sep=2pt, minimum height=4.2mm, font=\scriptsize},
  life/.style={draw=black!35, dash pattern=on 1.4pt off 1.4pt, line width=0.4pt},
  msg/.style={-{Latex[length=1.4mm]}, line width=0.5pt, qblue},
  qmsg/.style={-{Latex[length=1.4mm]}, line width=0.5pt, qred,
               dash pattern=on 2pt off 1.2pt},
  ml/.style={font=\tiny, text=black!75, inner sep=1pt, above, midway},
  act/.style={rounded corners=1.5pt, draw=qgreen, fill=qgreen!8, font=\tiny,
              align=left, inner sep=2pt, line width=0.4pt},
]
\def\ya{-0.55}\def\yb{-1.15}\def\yc{-1.75}\def\yd{-2.55}
\def\ye{-3.35}\def\yf{-3.95}\def\yg{-4.55}\def\yh{-5.35}

\node[actor] (C)  at (0,0)   {Controller};
\node[actor] (Kc) at (2.5,0) {KMS$_c$};
\node[actor] (Kv) at (5.0,0) {KMS$_v$};
\node[actor] (V)  at (7.5,0) {Verifier};

\foreach \n in {C,Kc,Kv,V} { \draw[life] (\n.south) -- ++(0,-5.9); }

\draw[msg] (C.south|-0,\ya) -- node[ml]{POST \texttt{enc\_keys}} (Kc.south|-0,\ya);
\node[font=\tiny, text=black!60, anchor=north] at (1.25,\ya-0.09)
  {\texttt{additional\_saes}, \texttt{number}, \texttt{size}};

\draw[qmsg] (Kc.south|-0,\yb) -- node[ml]{key relay (QKD)} (Kv.south|-0,\yb);

\draw[msg] (Kc.south|-0,\yc) -- node[ml]{\texttt{seed\_key}, \texttt{key\_id}} (C.south|-0,\yc);

\node[act, anchor=west] at (0.18,\yd)
  {\texttt{DGSP::keygen\_manager\_from\_seed}\\derives $gpk,\ gsk$; PLM, RL initialised};

\draw[msg] (C.south|-0,\ye) -- node[ml]{\texttt{SyncGroup}: \texttt{group\_name}, \texttt{key\_id}, relay pk} (V.south|-0,\ye);

\draw[msg] (V.south|-0,\yf) -- node[ml]{GET \texttt{dec\_keys}(\texttt{key\_id})} (Kv.south|-0,\yf);
\draw[msg] (Kv.south|-0,\yg) -- node[ml]{same \texttt{seed\_key}} (V.south|-0,\yg);

\node[act, anchor=east] at (7.35,\yh)
  {same deterministic derivation\\$\Rightarrow$ identical $gpk,\ gsk$ locally};

\end{tikzpicture}
\caption{On-demand group creation with deterministic QKD seeding. The group secret never traverses the classical channel: the controller and the verifier each retrieve the same seed  and re-derive identical group material.}
\label{fig:seeding}
\end{figure}

Deterministic seeding (Fig.~\ref{fig:seeding}) removes the central dependency 
from the message path: controller and verifiers feed the same key block, requested 
from their KME over ETSI GS QKD 014, as master seed to the scheme's derivation 
functions,  so every node derives \gpk{} and \gsk{} locally.

The controller requests a key from its trusted KME, naming through the multi-SAE 
authorisation  the verifiers that must obtain it; the central KME relays the 
entropy to their KMEs, consuming symmetric keys those pairs previously generated by QKD. 
From the returned \texttt{seed\_key} the controller derives the manager material, 
initialises the group state with PLM and RL empty, and propagates a \texttt{Sync Group}
message with the \texttt{group\_name}, the \texttt{key\_id} and the group's ML-DSA-65 
relay key pair, never \gsk{}; each verifier then retrieves the same seed from its own 
KME and repeats the derivation. Boot synchronisation, which would rebuild a restarted 
verifier's material from the stored \texttt{key\_id} of each active group, is design 
only and not exercised in the testbed.

Deterministic seeding hands every verifier \gsk{}. Signature opening, revocation 
and WOTS+ certification are nonetheless confined to the controller by policy, and only 
the controller holds the PLM binding a \texttt{user\_id} to a real vehicle. A 
compromised verifier could therefore open signatures and correlate those of one 
user, breaking that user's unlinkability, and could mint WOTS+ credentials with \gsk{}; 
without the PLM it still could not associate a \texttt{user\_id} with a real vehicle. 
Both boundaries rest on deployment policy, not cryptography, so the residual exposure 
of a verifier compromise is pseudonymous linkability rather than deanonymisation.

\section{Integration with ETSI ITS Messaging}
\label{sec:integ}

qDGSP keeps the DENM basic service of EN~302~637-3~\cite{etsi_en_302_637_3} and 
its GeoNetworking dissemination, adding a non-normative role container (\texttt{RolesContainer}) naming the originating and target functional group 
\emph{inside} the signed body, so neither can be rewritten without invalidating 
the signature. Baseline mode signs each DENM with a complete DGSP group signature 
$\sigma$.

\subsection{Registration and certificate provisioning}
\label{subsec:join}\label{subsec:csr}

Both exchanges need \gsk{} and therefore run against the controller over the
V2N core plane, off the DENM critical path. To register, the vehicle sends a
\texttt{JoinRequest} with its long-term credentials and functional group 
(\emph{Civil}, \emph{Emergency}, or another); the controller runs \texttt{Join}, 
yielding \usk{} and the logical identifier \texttt{user\_id}, inserts the entry 
in the PLM and pushes a \texttt{SyncJoinEvent} to the verifiers.

For provisioning, the OBU generates $B$ WOTS+ one-time key pairs locally and 
sends the public halves in a \texttt{CSR}; the controller checks the requester 
against the PLM, signs each WOTS+ public key with \gsk{} under SPHINCS+ (\texttt{GenCert}),
and returns a pool of $B$ certificates. Each certificate is consumed by exactly one group
signature, so the pool sizes the vehicle's anonymous emission budget. Provisioning is 
offline with respect to the critical path: it runs, with renewal, during idle periods 
or as low-priority asynchronous bursts, so the OBU always holds usable material 
without exposing its identity while driving.

\subsection{Message transit, role-based routing, and revocation cycle}
\label{subsec:transit}

The verifier applies four stages. (1) It checks the \texttt{ActionID} against a 
least-recently-used (LRU) cache with a 4\,s expiry time, discarding repetitions 
before any cryptographic work; the cache is shared by both signing modes, so an 
\texttt{ActionID} consumed by a baseline DENM cannot reappear as an EA intra-cell 
DENM, closing a cross-mode replay path. (2) The envelope's \texttt{sender\_group} 
and \texttt{target\_group} must match the signed role container; a mismatch is 
logged and dropped without triggering revocation. (3) \texttt{Verify} runs on $\sigma$ 
against \gpk{}. (4) The body's \texttt{CauseCode} must lie in the 
\texttt{allowed\_cause\_codes} of the role container's group: \emph{Civil} may 
emit $\{1, 2, 94, 99\}$ and \emph{Emergency} additionally $95$
(\textit{emergencyVehicleApproaching}), following~\cite{etsi_en_302_637_3}. 
A code outside the list is a semantic violation.

A message clearing all four stages is wrapped in an \texttt{AlertMessage} with 
the original DENM, the timestamps behind the T1--T6 components, a list that 
stops alerts looping on the backhaul, and a detached ML-DSA-65~\cite{nist_fips_204} 
relay signature under the target group's relay key. It propagates at once by 
V2I downlink to target-group OBUs in its own cell and by asynchronous HTTP/JSON 
backhaul to neighbouring verifiers. A semantic violation, or a dispute over an accepted message, opens the 
controlled-traceability cycle.

\section{qDGSP-EA: Ephemeral Intra-Cell Association}
\label{sec:ea}

The baseline mode pays a complete group signature per DENM: DGSP signatures 
inherit the size of the SPHINCS+ hypertree~\cite{sphincs+}, so one alert reaches 
tens of kilobytes and tens of IP fragments (Table~\ref{tab:size}), and one lost 
fragment destroys the datagram. Yet consecutive alerts from one vehicle in one 
cell re-prove the same anonymous membership in the same space--time window. 
qDGSP-EA therefore amortizes the group signature over a residence rather than 
over a message: on entering a cell the vehicle sends a signaling EA Announce, 
which pays the full DGSP cost and carries an ephemeral public key, and signs every
later alert in that cell with the matching ephemeral secret key. Unlike the 
credentials of~\cite{liu2023novelcomputationallyefficientgroup}, ours is anchored 
in a post-quantum group signature.

\subsection{Association and Intra-Cell Emission}

\begin{figure}[!t]
\centering
\begin{tikzpicture}[
  font=\scriptsize,
  actor/.style={rounded corners=2pt, draw=black!55, fill=black!4,
                inner sep=2.5pt, minimum height=4.2mm, font=\scriptsize},
  life/.style={draw=black!35, dash pattern=on 1.4pt off 1.4pt, line width=0.4pt},
  msg/.style={-{Latex[length=1.4mm]}, line width=0.5pt, qblue},
  lmsg/.style={-{Latex[length=1.4mm]}, line width=0.5pt, qorange},
  ml/.style={font=\tiny, text=black!75, inner sep=1pt, above, midway},
  act/.style={rounded corners=1.5pt, draw=qgreen, fill=qgreen!8, font=\tiny,
              align=left, inner sep=2pt, line width=0.4pt},
  phase/.style={font=\tiny\bfseries, text=black!60, anchor=west},
]
\node[actor] (O) at (0,0)   {OBU (vehicle)};
\node[actor] (V) at (6.6,0) {Verifier (Roadside ITS-S)};
\draw[life] (O.south) -- ++(0,-6.15);
\draw[life] (V.south) -- ++(0,-6.15);

\node[phase] at (-0.55,-0.45) {\textcircled{\tiny 1} identity anchoring};
\node[act, anchor=west] at (-0.5,-1.0)
  {generate ephemeral ML-DSA-65 pair\\(\texttt{ml\_pk\_lap}, \texttt{ml\_sk\_lap})};
\draw[msg] (0,-1.72) -- node[ml]{\texttt{Announce}\{group, \texttt{ml\_pk\_lap}\}, $\sigma_{\mathrm{ann}}$ (DGSP)} (6.6,-1.72);
\node[font=\tiny, text=qred, anchor=west] at (0.05,-2.02) {start $T_{\mathrm{ack}}=200$\,ms};

\node[phase] at (-0.55,-2.45) {\textcircled{\tiny 2} acknowledgement};
\node[act, anchor=east] at (6.45,-3.0)
  {verify $\sigma_{\mathrm{ann}}$ against $gpk$;\\record association in \texttt{obu\_registry}};
\draw[lmsg] (6.6,-3.6) -- node[ml]{\texttt{AnnounceAck} (ML-DSA-65 relay signature)} (0,-3.6);
\node[font=\tiny, text=black!60, anchor=north west] at (0.05,-3.72)
  {on timeout, retransmit; at most 3 attempts, then abort for this cell};

\node[phase] at (-0.55,-4.35) {\textcircled{\tiny 3} intra-cell emission};
\draw[msg] (0,-4.85) -- node[ml]{\texttt{DenmIntra}, ML-DSA-65 signature ($\times M$)} (6.6,-4.85);
\node[act, anchor=east] at (6.45,-5.4)
  {$O(1)$ registry lookup, verify,\\role check, forward};
\node[act, anchor=west, draw=qred, fill=qred!7] at (-0.5,-5.95)
  {on leaving the cell: destroy the ephemeral pair\\$\Rightarrow$ inter-cell unlinkability};

\end{tikzpicture}
\caption{qDGSP-EA association and intra-cell emission.}
\label{fig:ea}
\end{figure}
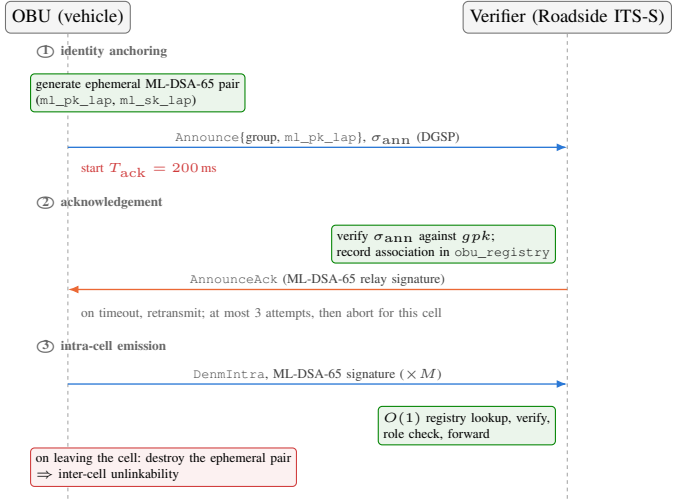

Fig.~\ref{fig:ea} shows the three phases.

\subsubsection{Identity anchoring} The OBU generates an ephemeral ML-DSA-65 key pair
and signs a length-prefixed encoding of its functional group and the ephemeral public 
key with a WOTS+ certificate from its DGSP pool. The resulting EA Announce carries 
that group signature \sigann{} on the fronthaul plane, and the OBU starts an 
application timer.

\subsubsection{EA Announce--ACK reliability} The verifier validates \sigann{} 
against \gpk{}, checks the declared group, registers the OBU in a 
volatile registry table binding each ephemeral key to the group signature, and returns 
an \texttt{Ack} signed with the ML-DSA-65 relay key of the OBU's own group, 
needing no extra key distribution. The OBU waits 200\,ms and retransmits 
on expiry, up to three transmissions; upon failure, it emits no intra-cell DENM, 
and the registry entry is simply never used.

\subsubsection{Intra-cell emission} Each later EA intra-cell DENM is signed with 
the ephemeral secret key alone and validated by one in-memory registry lookup and 
one ML-DSA-65 check instead of a DGSP verification, at the CPU cost of a pseudonym 
scheme. The alert then travels the backhaul as in baseline mode. The OBU destroys 
the ephemeral pair on leaving the cell.

\subsection{Privacy Properties and an Implementation Shortcut}

Inter-cell unlinkability is preserved: no ephemeral material crosses a cell
boundary and each EA Announce is itself an anonymous group signature. Intra-cell 
messages are linkable: an observer attributes every EA intra-cell DENM under one 
ephemeral key to one vehicle for its whole residence. EA mode thus trades the 
per-message unlinkability of baseline mode for per-residence unlinkability, bounded 
by the residence window. As a shortcut, the implementation indexes the OBU registry 
by the UDP  source address rather than by a cryptographic tag derived from the ephemeral 
public key.

Ephemeral keys change containment: the offending message carries no group s
ignature to open, so the response spans two horizons. Locally, a verifier 
detecting a semantic role violation on an EA intra-cell DENM, purges the 
ephemeral public key from the registry, so later EA intra-cell DENMs from that 
source are rejected. Globally and in parallel, the verifier sends the controller 
the stored \sigann{} of the EA Announce that established the purged key. 
Since \sigann{} identifies the root member cryptographically, the controller runs 
the \texttt{Open}--\texttt{Judge}--\texttt{Revoke} sequence, writes the offender 
to the RL and propagates it with a \texttt{SyncRevokeEvent}; the vehicle can 
then no longer produce a valid EA Announce anywhere in the deployment. 

\section{Evaluation}
\label{sec:eval}

The testbed runs every entity as a Docker Compose container over transport 
with realistic delay and loss
and measures the cryptographic cost between signing at the OBU and acceptance at 
the receiving vehicle. The application logic is Rust, extended only with 
\texttt{DGSP::keygen\_manager\_from\_seed}, which derives manager material from 
a QKD seed instead of the system random generator;  the SPHINCS+ parameter set is 
selected by feature flag (\texttt{128s}, \texttt{128f}, 
\texttt{256s}, \texttt{256f}), defaulting to \texttt{shake-256s} per 
FIPS~205, and experiments that do not sweep the parameter set 
use that default. The DKMS simulator exposes the ETSI GS QKD 014 endpoints, 
enforcing the requested multi-SAE authorisation; it reproduces the interface semantics 
of a QKD network, not its physics.

The verifier count is a Compose profile, $N\in\{1,3,5\}$; up to 500 logical vehicles 
share one container in the load runs. Each vehicle stays with a verifier for an 
exponentially distributed time of mean 60\,s, then jumps to another at random, 
following 5G NR V2X field trials and urban residence 
models~\cite{SANTA2022100537,8290681}; single-verifier load runs hold the 
association throughout. The four communication planes are Docker networks shaped 
with \texttt{netem} under the profiles of Table~\ref{tab:netem}, each the midpoint 
of its cited range, with normal jitter and independent Bernoulli loss per packet.

\begin{table}[!t]
\caption{Emulated Link Profiles; One-Way Delays. \texttt{qkd\_net} Is Unshaped.}
\label{tab:netem}
\centering
\begin{tabular}{lcccc}
\toprule
\textbf{Network} & \textbf{Delay} & \textbf{Jitter} & \textbf{Loss} & \textbf{Derived from} \\
\midrule
\texttt{fronthaul\_net} (V2I) & 10\,ms & 3\,ms & 1\,\% & \cite{SANTA2022100537,5G-MECTestbeds} \\
\texttt{core\_net} (V2N) & 25\,ms & 5\,ms & 0.5\,\% & \cite{Campolo,5G-MECTestbeds} \\
\texttt{backhaul\_net} & 5\,ms & 1\,ms & 0.1\,\% & \cite{5G-MECTestbeds} \\
\texttt{qkd\_net} & $<$1\,ms & --- & 0\,\% & \cite{etsi_qkd_014} \\
\bottomrule
\end{tabular}
\end{table}
.

\subsection{Cost}
\label{subsec:eval-cost}

\subsubsection{Wire size and fragmentation}
For a common 49\,B UPER body, the schemes span more than two orders of magnitude 
in wire bytes and IP fragments (Table~\ref{tab:size}). Fragmentation is the harsher 
metric: a datagram is lost whenever any one of its $k$ fragments is. Under the 
1\,\% per-packet loss of the \texttt{fronthaul\_net} profile, a single DGSP-256s 
transmission, spread over 23 fragments, fails with probability 
$1-0.99^{23}\approx20.6\,\%$; DGSP-256f, at 38 fragments, reaches 
$\approx31.7\,\%$. The EA intra-cell DENM occupies 3\,429\,B against 
3\,442\,B for ML-DSA-65, three fragments in both cases: on the intra-cell critical 
path qDGSP-EA costs no more than the NIST-standardized post-quantum 
reference~\cite{nist_fips_204}, with a single-transmission failure probability of
$\approx3.0\,\%$. The EA Announce reaches 34\,006\,B in 25 fragments, 
slightly larger than an isolated DGSP-256s DENM because its 1\,965\,B UPER body 
embeds the 1\,952\,B ephemeral ML-DSA-65 public key, a price paid once per cell.

\begin{table}[!t]
\renewcommand{\arraystretch}{1.01}
\caption{Wire Size and IP Fragmentation per Scheme. ``Wire'' is the complete UDP payload, envelope and signature included; fragments are counted in 1\,400\,B effective MTUs; ``Ratio'' is relative to ECDSA P-256.}
\label{tab:size}
\centering
\begin{tabular}{lcccc}
\toprule
\textbf{Scheme} & \textbf{UPER (B)} & \textbf{Wire (B)} & \textbf{Frag.} & \textbf{Ratio} \\
\midrule
ECDSA P-256          & 49 &      197 &  1 &   1.0$\times$ \\
ML-DSA-65            & 49 & 3\,442  &  3 &  17.5$\times$ \\
Falcon-512           & 49 &      788 &  1 &   4.0$\times$ \\
DGSP-128s            & 49 & 8\,576  &  7 &  43.5$\times$ \\
DGSP-128f            & 49 & 17\,808 & 13 &  90.4$\times$ \\
DGSP-256s (default)  & 49 & 32\,128 & 23 & 163.1$\times$ \\
DGSP-256f            & 49 & 52\,192 & 38 & 264.9$\times$ \\
\midrule
EA Announce          & 1\,965 & 34\,006 & 25 & 172.6$\times$ \\
EA intra-cell DENM   & 49 & 3\,429  &  3 &  17.4$\times$ \\
\bottomrule
\end{tabular}
\end{table}

\subsubsection{Signing and verification CPU}
Signing DGSP-256s costs 4\,ms in median, against 2\,ms for ML-DSA-65 and 21\,ms 
for Falcon-512. Verification is where DGSP is expensive: 11\,ms with the default 
parameter set (p95 13\,ms) and 22\,ms with the fast-signing variant DGSP-256f, 
a cost borne by the verifier once per message and per verifier. The EA intra-cell 
DENM verifies below one millisecond, as the ML-DSA-65 reference does.

\subsubsection{Provisioning}
Sweeping the WOTS+ pool size $B$ over $\{1, 8, 32, 100\}$, both the vehicle-side 
key generation and the total CSR cost grow approximately linearly in $B$ once the 
per-certificate SPHINCS+ signing dominates the fixed overhead: between $B=1$ and $B=8$ 
the total is nearly flat, 3.0\,s against 3.4\,s, because the V2N round trip dominates, 
while a pool of $B=32$ costs 6.6\,s in median and $B=100$ reaches 19.7\,s. No V2X flow 
on the critical path tolerates that cost.

\subsubsection{qDGSP-EA amortization}
Table~\ref{tab:amort} compares the aggregate per-cell cost of the two modes for 
$M\in\{1,3,5,10\}$ alerts per cell. At $M=1$ EA mode is the more expensive option, 
the EA Announce being paid without being shared; break-even lies between $M=1$ and 
$M=3$, and the saving grows monotonically to $+78.7\,\%$ in bytes and $+76.0\,\%$ 
in CPU at $M=10$. Bytes and CPU amortize at nearly the same rate, both being 
dominated by SPHINCS+ signatures.

\begin{table}[!t]
\caption{Aggregate Per-Cell Cost of qDGSP-EA Against Baseline Mode ($M$ DGSP-Signed DENMs).}
\label{tab:amort}
\centering
\begin{tabular}{ccccc}
\toprule
$M$ & \textbf{EA (KB)} & \textbf{Baseline (KB)} & \textbf{Byte saving} & \textbf{CPU saving} \\
\midrule
1  & 36.56 & 31.38  & $-16.5\,\%$ & $-20.0\,\%$ \\
3  & 43.25 & 94.12  & $+54.0\,\%$ & $+51.1\,\%$ \\
5  & 49.95 & 156.88 & $+68.2\,\%$ & $+65.3\,\%$ \\
10 & 66.70 & 313.75 & $+78.7\,\%$ & $+76.0\,\%$ \\
\bottomrule
\end{tabular}
\end{table}

\subsection{Operational Feasibility}
\label{subsec:eval-feas}

\begin{figure}[!t]
  \centering
  \includegraphics[width=0.49\columnwidth]{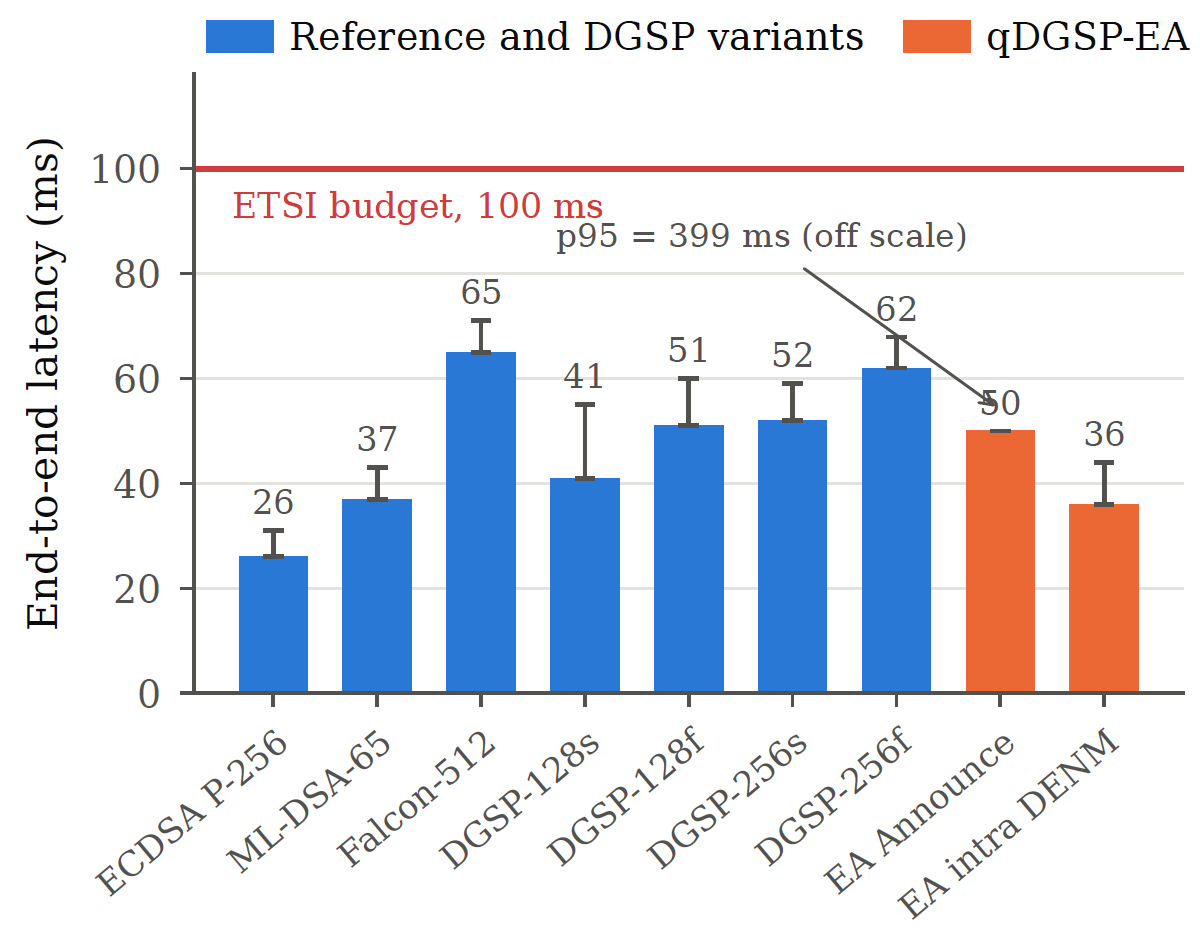}
  \includegraphics[width=0.49\columnwidth]{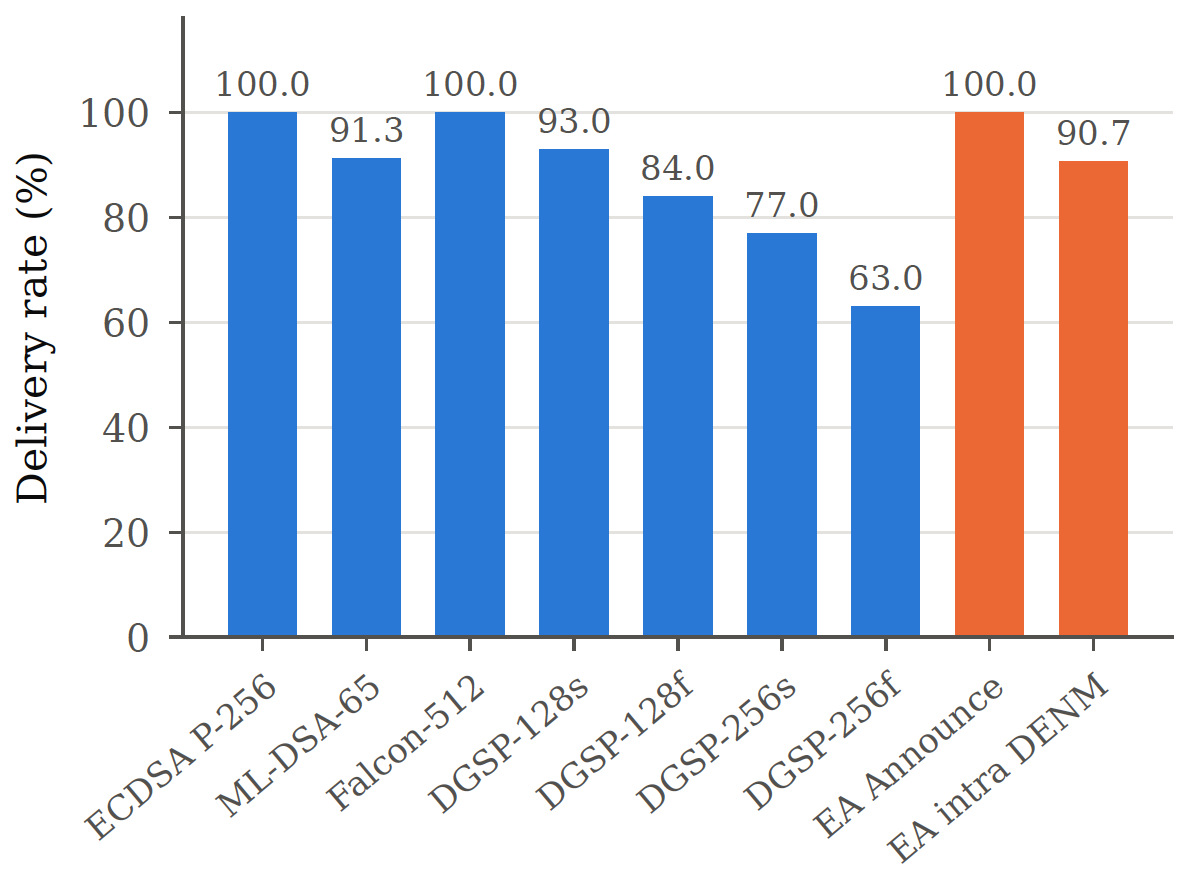}
  \caption{End-to-end latency and delivery rate per scheme, at one transmission per DENM and 1\,\% fronthaul loss.}
  \label{fig:e2e}
\end{figure}

\subsubsection{End-to-end latency and delivery per scheme}
Fig.~\ref{fig:e2e} reports median end-to-end (E2E) latency, its p95, and the 
rate of unique DENMs delivered to the Emergency OBU within the 100\,ms budget, 
for the nine schemes at \texttt{DENM\_REPETITIONS=1} and 1\,\% fronthaul loss. 
Every median meets the ETSI budget: the slowest scheme is Falcon-512 at 65\,ms, 
penalized by its signing cost, followed by the most expensive DGSP variant, DGSP-256f, 
at 62\,ms with a p95 of 67.9\,ms. Conditioned on the DENM arriving at all, it arrives 
within budget in at least 95\,\% of cases for the eight DENM-carrying schemes.
Delivery falls from 100.0\,\% for the single-fragment schemes to 77.0\,\% 
for DGSP-256s over 23 fragments and 63.0\,\% for DGSP-256f over 38. 
The EA Announce records 100.0\,\% delivery over the 26 samples of this experiment, 
with a median of 50\,ms yet a p95 of 399\,ms.

\begin{figure}[!t]
  \centering
  \includegraphics[width=\columnwidth]{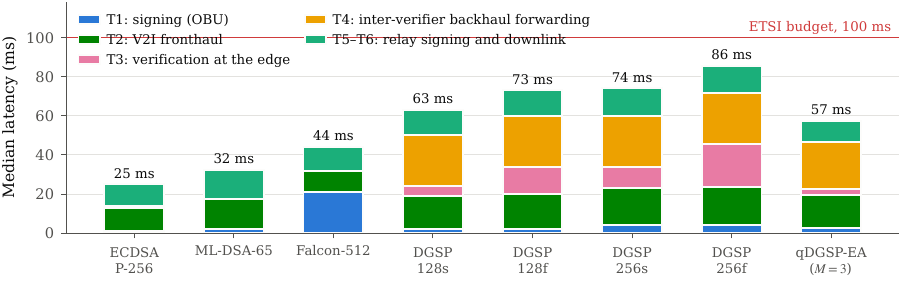}
  \caption{Median end-to-end latency split into its components.  Each bar sums per-component medians.}
  \label{fig:decomp}
\end{figure}

\subsubsection{Latency decomposition}
Fig.~\ref{fig:decomp} splits the median E2E latency into signing at the 
emitting OBU (T1), V2I fronthaul transit until the last IP fragment reaches 
the verifier (T2), edge verification (T3), asynchronous forwarding of the \texttt{AlertMessage} to neighboring verifiers (T4), and the relay verifier's 
signature, the downlink hop and decoding at the receiving OBU (T5--T6). 
T4 reaches 26\,ms for DGSP-256s at $N=3$ and is zero for the reference schemes; 
the EA Announce produces no T4 either, since the Announce--ACK exchange closes between 
the OBU and its local verifier. T2 scales with fragmentation, lowest for the 
single-fragment schemes and highest for the heavy DGSP variants.

\begin{figure}[!t]
  \centering
  \includegraphics[width=0.8\columnwidth]{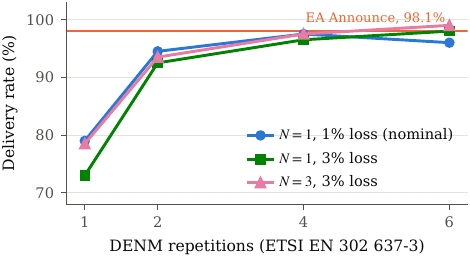}
  \caption{DENM delivery rate against the number of ETSI repetitions, under the 1\,\% and 3\,\% V2I loss profiles.}
  \label{fig:reps}
\end{figure}

\subsubsection{Delivery against ETSI repetitions}
Fig.~\ref{fig:reps} covers three loss and topology configurations. With a single 
transmission, delivery lies between 73.0\,\% and 79.0\,\%; it improves asymptotically 
with the repetition count and from four to six repetitions exceeds 95\,\% on every
configuration. The repetition mechanism
therefore absorbs the fragmentation penalty at the price of channel occupancy proportional 
to message size. The EA Announce 
is measured at 98.1\,\%: a few associations exhaust the three attempts without a reply, 
after which the OBU skips association to that cell.

\subsubsection{Behavior under concurrent load}
The load experiment sweeps $V\in\{50,100,500\}$ concurrent logical vehicles 
against $N\in\{1,3\}$ verifiers, the two lower levels at the densities of the 
V2AIX capture (285\,312 ETSI ITS messages, 2\,388 stations, 30.07\,h)~\cite{v2aixmultimodalrealworlddataset} and $V=500$ deliberately beyond 
the realistic range for a single verifier, probing induced saturation.
The median is flat in $V$: baseline mode at $N=1$ measures 40, 40 and 39\,ms as $V$ 
goes from 50 to 500, EA mode at $N=1$ stays at 31\,ms, and the p5--p95 band does not 
widen appreciably, so neither the verifier's CPU nor the fronthaul input queue is the
bottleneck in the evaluated range. The penalty comes from $N$: moving from one to 
three verifiers adds up to 19\,ms to the median, dominated by component T4 rather 
than concurrency.

\subsubsection{Revocation propagation}
With $N=5$, the five verifiers apply a confirmed revocation at 10, 30, 52, 73 and 97\,ms 
from the close of the cycle. Growth is approximately linear in $N$, at roughly 20\,ms per
additional verifier, set by the serialized HTTP POST that the controller issues to each
verifier inside its dispatch loop plus the backhaul link profile of Table~\ref{tab:netem}.
Complete propagation at $N=5$ therefore falls in the same order as the end-to-end budget
of a single DENM, and this term grows with verifier density.

\balance
\bibliographystyle{IEEEtran}
\bibliography{refs}

@misc{dgsp2025,
  title  = {{DGSP}: An Efficient Scalable Fully Dynamic Group Signature Scheme Using {SPHINCS+}},
  author = {Mojtaba Fadavi and Seyyed Arash Azimi and Sabyasachi Karati and Samuel Jaques},
  year   = 2025,
  note   = {\url{https://eprint.iacr.org/2025/760}}
}

@inproceedings{GroupSignature,
  title     = {Group Signatures},
  author    = {Chaum, David and van Heyst, Eug{\`e}ne},
  year      = 1991,
  booktitle = {Advances in Cryptology --- EUROCRYPT '91},
  pages     = {257--265},
  publisher = {Springer Berlin Heidelberg}
}

@inproceedings{sphincs+,
  title     = {The SPHINCS+ Signature Framework},
  author    = {Bernstein, Daniel J. and H\"{u}lsing, Andreas and K\"{o}lbl, Stefan and Niederhagen, Ruben and Rijneveld, Joost and Schwabe, Peter},
  year      = 2019,
  booktitle = {Proceedings of the 2019 ACM SIGSAC Conference on Computer and Communications Security},
  pages     = {2129–2146},
  doi       = {10.1145/3319535.3363229}
}

@inproceedings{wots,
  title     = {W-OTS+ -- Shorter Signatures for Hash-Based Signature Schemes},
  author    = {H{\"u}lsing, Andreas},
  year      = 2013,
  booktitle = {Progress in Cryptology -- AFRICACRYPT 2013},
  pages     = {173--188},
  publisher = {Springer Berlin Heidelberg}
}

@techreport{nist_fips_204,
  author      = {{National Institute of Standards and Technology}},
  title       = {{Module-Lattice-Based Digital Signature Standard}},
  institution = {U.S. Department of Commerce},
  type        = {FIPS},
  number      = {NIST FIPS 204},
  year        = {2024},
  doi         = {10.6028/NIST.FIPS.204}
}

@techreport{etsi_tr_102_638,
  author      = {{ETSI}},
  title       = {{TR 102 638 V1.1.1: Intelligent Transport Systems (ITS); Vehicular Communications; Basic Set of Applications; Definitions}},
  institution = {ETSI},
  year        = {2009}
}

@techreport{etsi_ts_122_186_v19,
  author      = {{ETSI}},
  title       = {{TS 122 186 V19.0.0: {5G}; Service requirements for enhanced {V2X} scenarios}},
  institution = {ETSI},
  year        = {2025}
}

@misc{dervisevic2024quantumkeydistributionnetworks,
      title={Quantum Key Distribution Networks -- Key Management: A Survey}, 
      author={Emir Dervisevic and Amina Tankovic and Ehsan Fazel and Ramana Kompella and Peppino Fazio and Miroslav Voznak and Miralem Mehic},
      year={2024},
      eprint={2408.04580},
      archivePrefix={arXiv},
      primaryClass={cs.CR}
}

@techreport{etsi_qkd_014,
  author      = {{ETSI}},
  title       = {{GS QKD 014 V1.1.1: Quantum Key Distribution (QKD); Protocol and data format}},
  institution = {ETSI},
  year        = {2019},
  month       = {February}
}

@techreport{ETSI_EN_302_665_V1.1.1,
  author      = {{ETSI}},
  title       = {{Intelligent Transport Systems (ITS); Communications Architecture}},
  institution = {European Telecommunications Standards Institute},
  number      = {ETSI EN 302 665 V1.1.1},
  year        = {2010}
}

@techreport{etsi_en_302_637_3,
  author      = {{ETSI}},
  title       = {{ETSI EN 302 637-3 V1.3.1: ITS; Vehicular Communications; Decentralized Environmental Notification Basic Service}},
  institution = {ETSI},
  year        = {2019}
}

@techreport{etsi_ts_103097_v211,
  author      = {{ETSI}},
  title       = {{Intelligent Transport Systems (ITS); Security; Security header and certificate formats}},
  institution = {ETSI},
  number      = {ETSI TS 103 097 V2.1.1},
  year        = {2021}
}

@techreport{etsi_en_302636_5_1,
  author      = {{ETSI}},
  title       = {{Intelligent Transport Systems (ITS); Vehicular Communications; GeoNetworking; Part 5: Transport Protocols; Sub-part 1: Basic Transport Protocol}},
  institution = {{European Telecommunications Standards Institute (ETSI)}},
  type        = {{ETSI EN}},
  number      = {302 636-5-1 V2.2.1},
  month       = {05},
  year        = {2019}
}

@misc{liu2023novelcomputationallyefficientgroup,
  title  = {A Novel Computationally Efficient Group Signature for Anonymous and Secure {V2X} Communications}, 
  author = {Jia Liu and Liqun Chen and Mehrdad Dianati and Carsten Maple and Yan Yan},
  year   = {2023},
  eprint = {2307.13759},
  archivePrefix = {arXiv}
}

@article{5G-MECTestbeds,
  author  = {Wadatkar, Prachi V. and Garroppo, Rosario G. and Nencioni, Gianfranco},
  title   = {{5G-MEC} Testbeds for {V2X} Applications},
  journal = {Future Internet},
  volume  = {15},
  year    = {2023},
  number  = {5},
  doi     = {10.3390/fi15050175}
}

@article{8290681,
  author  = {Aissioui, Abdelkader and Ksentini, Adlen and Gueroui, Abdelhak Mourad and Taleb, Tarik},
  journal = {IEEE Trans. on Vehicular Technology}, 
  title   = {On Enabling {5G} Automotive Systems Using Follow Me Edge-Cloud Concept}, 
  year    = {2018},
  volume  = {67},
  number  = {6},
  pages   = {5302-5316},
  doi     = {10.1109/TVT.2018.2805369}
}

@inproceedings{Campolo,
  author    = {Campolo, Claudia and Iera, Antonio and Molinaro, Antonella and Ruggeri, Giuseppe},
  booktitle = {2019 IEEE Wireless Communications and Networking Conference (WCNC)}, 
  title     = {MEC Support for {5G-V2X} Use Cases through Docker Containers}, 
  year      = {2019},
  pages     = {1-6},
  doi       = {10.1109/WCNC.2019.8885515}
}

@misc{v2aixmultimodalrealworlddataset,
  title  = {V2AIX: A Multi-Modal Real-World Dataset of ETSI ITS {V2X} Messages in Public Road Traffic}, 
  author = {Guido Kueppers and Jean-Pierre Busch and Lennart Reiher and Lutz Eckstein},
  year   = {2024},
  eprint = {2403.10221},
  archivePrefix = {arXiv}
}

@article{SANTA2022100537,
  title   = {Evaluation platform for {5G} vehicular communications},
  journal = {Vehicular Communications},
  volume  = {38},
  year    = {2022},
  author  = {Jose Santa and Konstantinos V. Katsaros and Luis Bernal-Escobedo and others},
  doi     = {10.1016/j.vehcom.2022.100537}
}

@inproceedings{netem,
  author    = {Hemminger, Stephen},
  title     = {Network Emulation with {NetEm}},
  booktitle = {Linux Conference Australia (LCA)},
  year      = {2005},
  pages     = {18--23},
  address   = {Canberra, Australia}
}

@misc{techrxiv_pqc_v2x,
  author       = {Twardokus, Geoff and Rahbari, Hanif},
  title        = {Quantum-Resistant Safety Message Authentication for {NextG} {C-V2X}: A Cross-Layer Approach},
  year         = {2025},
  howpublished = {TechRxiv preprint},
  doi          = {10.36227/techrxiv.175751197.77663474}
}

@article{Sedar_V2X_security_survey,
  author  = {Sedar, Roshan and Kalalas, Charalampos and Vázquez-Gallego, Francisco and Alonso, Luis and Alonso-Zarate, Jesus},
  title   = {A Comprehensive Survey of {V2X} Cybersecurity Mechanisms and Future Research Paths},
  journal = {IEEE Open J. of the Commun. Society},
  volume  = {4},
  pages   = {325--391},
  year    = {2023},
  doi     = {10.1109/OJCOMS.2023.3239115}
}

@techreport{etsi_ts_102_941,
  author      = {{ETSI}},
  title       = {{TS 102 941 V2.2.1: Intelligent Transport Systems (ITS); Security; Trust and Privacy Management; Release 2}},
  institution = {ETSI},
  year        = {2022},
  month       = {November}
}

@article{perera_survey_groupring,
  author        = {Perera, Maharage Nisansala Sevwandi and Nakamura, Toru and Hashimoto, Masayuki and Yokoyama, Hiroyuki and Cheng, Chen-Mou and Sakurai, Kouichi},
  title         = {{A Survey on Group Signatures and Ring Signatures: Traceability vs. Anonymity}},
  journal       = {Cryptography},
  volume        = {6},
  number        = {1},
  pages         = {3},
  year          = {2022},
  doi           = {10.3390/cryptography6010003}
}

@article{kaaniche2020privacy,
  author  = {Kaaniche, Nesrine and Laurent, Maryline and Belguith, Sana},
  title   = {Privacy enhancing technologies for solving the privacy-personalization paradox: Taxonomy and survey},
  journal = {J. of Network and Computer App.},
  volume  = {171},
  pages   = {102807},
  year    = {2020},
  doi     = {10.1016/j.jnca.2020.102807}
}

\end{document}